\documentclass[a4paper,11pt]{article}
\usepackage{jheppub} % for details on the use of the package, please see the JINST-author-manual
\usepackage{lineno}
\arxivnumber{2509.17487} % if you have one

\title{Measurement of transverse polarization of {\boldmath$\Lambda$} and \boldmath{$\overline{\Lambda}$} hyperons inside jets in unpolarized proton-proton collisions at $\sqrt{s}$ = 200 GeV}

\author[59]{B.~E.~Aboona}
\author[17]{J.~Adam} 
\author[32]{G.~Agakishiev} 
\author[45]{I.~Aggarwal} 
\author[45]{M.~M.~Aggarwal} 
\author[66]{Z.~Ahammed} 
\author[32]{A.~Aitbayev} 
\author[3 41]{I.~Alekseev} 
\author[41]{E.~Alpatov} 
\author[33]{A.~K.~Alshammri} 
% ...

\affiliation{\rm{(STAR Collaboration)}}
% ...
\affiliation[3]{Alikhanov Institute for Theoretical and Experimental Physics NRC "Kurchatov Institute", Moscow 117218}
% ...
\affiliation[17]{Czech Technical University in Prague, FNSPE, Prague 115 19, Czech Republic}
% ...
\affiliation[32]{Joint Institute for Nuclear Research, Dubna 141 980}
\affiliation[33]{Kent State University, Kent, Ohio 44242}
% ...
\affiliation[41]{National Research Nuclear University MEPhI, Moscow 115409}
% ...
\affiliation[45]{Panjab University, Chandigarh 160014, India}
% ...
\affiliation[59]{Texas A\&M University, College Station, Texas 77843}
% ...
\affiliation[66]{Variable Energy Cyclotron Centre, Kolkata 700064, India}
% ...

\emailAdd{star-publication@bnl.gov}

\abstract{A surprisingly large transverse polarization of $\Lambda$ hyperons in unpolarized hadron-nucleon/nucleus collisions has been observed for 50 years, and the origin of this polarization remains an important open question.
Recently, theoretical frameworks have advanced in describing this puzzle with the polarizing fragmentation function (PFF).
We report the first measurement of $\Lambda$ and $\overline{\Lambda}$ transverse polarization inside jets in unpolarized proton-proton collisions, which is directly attributed to the PFF. The polarization is measured as a function of the jet transverse momentum, the fraction of the jet momentum carried by $\Lambda$($\overline{\Lambda}$) hyperons, and the transverse momentum of $\Lambda(\overline{\Lambda})$ hyperons relative to the jet axis. 
Covering a wide jet-energy range, these data provide the first constraints on the gluon PFF and allow tests of TMD evolution and its universality.
}

\begin{document}
\maketitle
\flushbottom

\section{Introduction} \label{sec:int}

%%%%%%%%%%%%%
Quantum chromodynamics (QCD), the theory of the strong interaction, exhibits asymptotic freedom at short distances and color confinement at long distances. While perturbative QCD successfully describes hard scattering processes, the formation of colorless hadrons from quarks and gluons is governed by nonperturbative dynamics and is not yet fully understood. Spin-dependent fragmentation dynamics of final-state hadrons provide a sensitive probe of this hadronization process. A striking manifestation of such dynamics is the large transverse polarization of $\Lambda$ hyperons produced in unpolarized hadronic collisions, first observed in 1976~\cite{Bunce:1976yb}. The polarization can reach values as large as $\sim 40\%$ in certain kinematic regions and has been observed across a wide range of collision systems and energies, including fixed-target experiments~\cite{Panagiotou:1989sv,LHCb:2024vwi}, neutrino scattering~\cite{NOMAD:2000wdf,NOMAD:2001iup}, lepton--nucleon scattering~\cite{HERMES:2007fpi}, $e^+e^-$ annihilation~\cite{ALEPH:1996oew,1998OPAL,Belle:2018ttu}, and hadron--hadron collisions up to the highest available energies~\cite{Gao:2024dxl,ATLAS:2014ona,STAR:2025njp,LHCb:2025Lambda}.
Perturbative QCD calculations predict negligible contributions to this observable~\cite{pQCD_cal}, indicating that transverse $\Lambda$ polarization originates from nonperturbative mechanisms associated with hadronization. Despite extensive experimental measurements and theoretical developments over the past decades~\cite{Felix:1999tf,Liang:1997rt,Boer:1997nt,Anselmino:2000vs,Anselmino:2001js,Kanazawa:2000cx,Zhou:2008fb,Boer:2007nh,Koike:2017fxr,Kang:2020xyq,Kang:2021kpt,Chen:2021zrr,Koike:2022ddx,Ikarashi:2022zeo,DAlesio:2023ozw,Ji:2023cdh,DAlesio:2024ope,Gao:2024bfp}, a complete and unified description of the underlying mechanism is still missing.

A modern theoretical framework for transverse $\Lambda$ polarization is provided by the polarizing fragmentation function (PFF)~\cite{Anselmino:2000vs,Boer:2007nh,Kang:2020xyq}, which describes the production of a transversely polarized hadron from the fragmentation of an unpolarized parton~\cite{Mulders:1995dh}. The PFF is one of the eight leading-twist transverse-momentum-dependent (TMD) fragmentation functions~\cite{Boussarie:2023izj}. Key open questions include its flavor dependence, the role of gluon fragmentation, and its universality across different processes. As a time-reversal-odd fragmentation function, the PFF is expected to be universal, in contrast to the Sivers parton distribution function, which exhibits a process-dependent sign change between semi-inclusive deep-inelastic scattering and the Drell--Yan process~\cite{Metz:2002iz,Collins:2004nx,Meissner:2008yf,Boer:2010ya}. Experimental tests of PFF universality therefore provide a qualitatively distinct probe of QCD dynamics compared to T-odd distribution functions.
Electron--positron annihilation offers a clean environment to access the quark PFF through measurements of $\Lambda$ polarization relative to the thrust axis. Early measurements at LEP at $\sqrt{s}=90$~GeV did not observe statistically significant polarization~\cite{ALEPH:1996oew,1998OPAL}. More recently, significant transverse polarization of $\Lambda(\overline{\Lambda})$ hyperons was observed at $\sqrt{s}=10.6$~GeV by the BELLE experiment~\cite{Belle:2018ttu}, enabling phenomenological extractions of the quark PFF~\cite{DAlesio:2020wjq,Callos:2020qtu,Chen:2021hdn,Li:2020oto,Gamberg:2021iat,DAlesio:2022brl}. However, $e^+e^-$ data are primarily sensitive to quark fragmentation and provide limited constraints on the gluon PFF.

Proton--proton collisions provide a complementary environment in which both quark and gluon fragmentation contribute. In particular, measurements of transverse $\Lambda(\overline{\Lambda})$ polarization inside jets give direct access to the PFF in jet fragmentation~\cite{Boer:2007nh,Kang:2020xyq,Kang:2017glf}. At RHIC energies, gluon-initiated subprocesses contribute substantially to inclusive jet production~\cite{DAlesio:2024ope,_2019_LongitudinalDoublespin}, making such measurements sensitive to the gluon PFF, which is not constrained by $e^+e^-$ data. In addition, the quark flavor composition of jets in $pp$ collisions differs significantly from that in $e^+e^-$ annihilation and varies with jet kinematics. Measurements in $pp$ collisions therefore provide sensitivity not only to gluon fragmentation, but also to the flavor dependence of quark PFFs beyond the flavor mixture accessible in $e^+e^-$ processes.

\begin{figure}[bht]
    \centering
\includegraphics[width=0.6\linewidth]{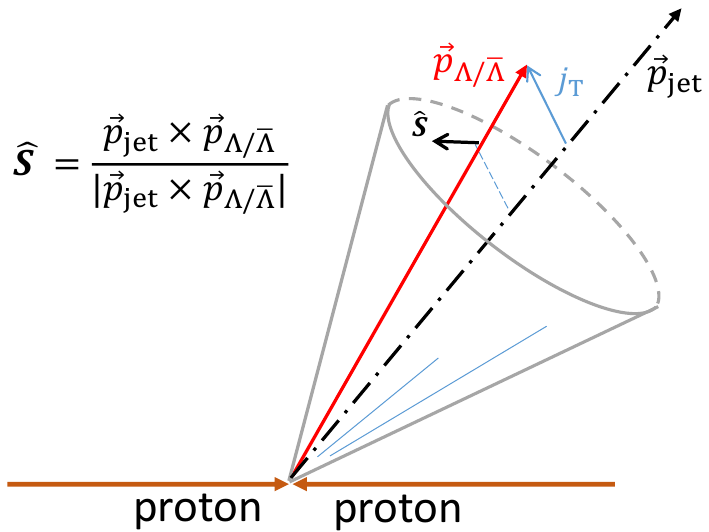}
	\caption{Schematic of $\Lambda(\overline{\Lambda})$ inside a jet in $pp$ collisions. The polarization direction $\hat{\boldsymbol{S}}$ is defined by the jet and $\Lambda$ momenta. \label{fig:Linjet}}
\end{figure}

In this paper, we present the first measurement of the transverse polarization of $\Lambda(\overline{\Lambda})$ hyperons inside jets in unpolarized $pp$ collisions at $\sqrt{s}=200$ GeV at the Relativistic Heavy-Ion Collider (RHIC) at Brookhaven National Laboratory. %Dataset used for this measurement were collected by the STAR experiment in 2015, with an integrated luminosity of 133 pb$^{-1}$.  
As shown in figure{}~\ref{fig:Linjet}, the polarization direction for $\Lambda(\overline{\Lambda})$ hyperons in this analysis is defined along the normal to the  $\Lambda(\overline{\Lambda})$ production plane inside a jet, denoted as $\hat{\boldsymbol{S}} = \frac{\vec{\boldsymbol{p}}_{\text{jet}} \times \vec{\boldsymbol{p}}_{\Lambda/\overline{\Lambda}}}{|\vec{\boldsymbol{p}}_{\text{jet}} \times \vec{\boldsymbol{p}}_{\Lambda/\overline{\Lambda}}|}$, with $\vec{\boldsymbol{p}}_{\text{jet}}$ and $\vec{\boldsymbol{p}}_{\Lambda/\overline{\Lambda}}$ being the jet and $\Lambda(\overline{\Lambda})$ momentum, respectively.
The $\Lambda(\overline{\Lambda})$ polarizations are measured as functions of the jet transverse momentum $p_{\text{T}}$, longitudinal momentum fraction $z$ of the jet carried by $\Lambda(\overline{\Lambda})$, and transverse momentum $j_{\text{T}}$ of $\Lambda(\overline{\Lambda})$ relative to the jet axis.

\section{Experiment and data analysis}
The dataset used for this measurement was collected by the STAR experiment at RHIC in 2015 using $pp$ collisions at $\sqrt{s}=200$ GeV, with an integrated luminosity of 133 pb$^{-1}$. 
%The proton spin direction alternates from bunch to bunch about every 106 ns at RHIC, but for this analysis, 
The proton beams are effectively unpolarized after summing over different beam spin configurations. 
Subsystems of the STAR detector~\cite{STAR_det} involved in this measurement are the Time Projection Chamber (TPC)~\cite{TPC}, the Barrel Electromagnetic Calorimeter (BEMC)~\cite{BEMC}, and the Endcap Electromagnetic Calorimeter (EEMC)~\cite{EEMC}. Embedded in a 0.5 T magnetic field, the TPC provides charged particle track reconstruction and identification for pseudorapidity $|\eta|\lesssim1.3$. The BEMC and EEMC are both lead-scintillator sampling calorimeters that cover the full azimuthal angle $\phi$ for pseudorapidity $|\eta|<1.0$ and $1.086 < \eta < 2$, respectively. 

The events used in this analysis are selected with jet patch (JP) triggers, which require the transverse electromagnetic energy ($E_{\text{T}}$) in a region $\Delta \eta$ $\times$ $\Delta \phi$ = 1.0 $\times$ 1.0 in the BEMC or EEMC to exceed a given threshold~\cite{star:2012jet}. 
%In this measurement, only events satisfying the BEMC and EEMC jet-patch trigger conditions (JP1 and JP2) are used. The jet-patch triggers at STAR has been described in detail in reference ~\cite{star:2012jet}.  
The transverse energy thresholds for JP1 and JP2 triggers are set to $5.4 $ GeV and $7.3$ GeV, respectively, in 2015. 
The event vertices are reconstructed from TPC tracks, and the $z$ component of the collision vertex is required to be within 90 cm (along the beamline) of the center of the TPC for uniform acceptance.
%The primary vertex position of each event along the beam line, $V_z$, is required to be within $\pm 90$ cm from the center of STAR detector. 

$\Lambda(\overline{\Lambda})$ candidates are reconstructed via
the dominant weak decay channel $\Lambda \rightarrow p + \pi^{-}$ ($\overline{\Lambda}\rightarrow \overline{p} + \pi^+$) from TPC tracks, following a similar procedure as was used in previous STAR measurements~\cite{star:2009dll, star:2018dll, star:2018dtt, star:2024dlltt, Chen:2026gka}.
(Anti-)Proton and pion tracks are identified based on their energy loss d$E$/d$x$ in the TPC, and a minimum transverse momentum of $0.15$ GeV$/c$ is required. (Anti-)Proton and pion tracks are then paired to form a $\Lambda(\overline{\Lambda})$ candidate, and a set of $p_{\text{T}}$-dependent topological selections is applied to suppress the combinatorial background. 
%The distance of closest approach (DCA) between daughter tracks is required to be small, typically less than 0.65$\sim$0.40 cm as $p_T$ increases. Besides, the DCA of daughter tracks to the primary vertex (PV) is set a minimum value decreasing as $p_T$ increasing. The reconstructed $\Lambda(\overline{\Lambda})$ candidate is required to have a small DCA to the PV, typically below 0.55 $\sim$ 1.0 cm. Additional selections are imposed on the decay length, which is required to be larger than about 3.0$\sim$4.5 cm, and on the pointing angle, with $\cos(\vec{r},\vec{p}) > 0.995$, where $\vec{r}, \vec{p}$ represent the vector from the primary event vertex to the decay point and momentum of $\Lambda(\overline{\Lambda})$. 
The distance of closest approach (DCA) between the daughter tracks is required to be small, decreasing from 0.65 cm to 0.40 cm as the hyperon $p_{\mathrm{T}}$ increases. Additionally, the DCA of the daughter tracks to the primary vertex (PV) is required to exceed a $p_{\mathrm{T}}$-dependent threshold. For the reconstructed $\Lambda(\overline{\Lambda})$ candidate, an upper limit on the DCA to the PV, which ranges from 0.55 cm to 1.0 cm as $p_{\mathrm{T}}$ varies. Further selections are applied to the decay length, which must be larger than a value ranging from approximately 3.0 cm to 4.5 cm as $p_{\mathrm{T}}$ increases, and to the pointing angle, requiring $\cos(\vec{r},\vec{p}) > 0.995$, where $\vec{r}$ and $\vec{p}$ denote the vector from the primary vertex to the decay point and the momentum of the $\Lambda(\overline{\Lambda})$, respectively. These criteria effectively suppress the combinatorial background while maintaining good signal efficiency.
After the selection, the purity of the $\Lambda(\overline{\Lambda})$ candidates is about $90\%$ in the mass peak region ($1.112 < M < 1.120 $ GeV/$c^2$). 
%Tracks are required to have at least 15 hit points and the ratio of hits over maximum possible hits to be larger than 0.52. The minimum transverse momentum is $0.2$ GeV$/c$. (Anti-)Proton and pion tracks are identified based on the energy loss per unit length d$E$/d$x$ in TPC by requiring $n|\sigma_{p(\pi)}|< 3.0$ where n$\sigma_{p(\pi)}$ denotes the deviation between the measured energy loss to the theoretical values. Candidate (anti-)proton and pion tracks are paired to form $\Lambda(\overline{\Lambda})$ candidate.  To suppress the background, a set of $p_{\text{T}}$ dependent topological criteria are applied, following the same procedure as used in previous measurements~\cite{star:2009dll, star:2018dll, star:2018dtt, star:2024dlltt}. 

Jets are reconstructed using the anti-$k_{\text{T}}$ algorithm~\cite{antiKt:2008gp} with a radius parameter $R=0.6$. The input particle list includes reconstructed $\Lambda(\overline{\Lambda})$ candidates, primary charged tracks from the TPC, and energy deposits in the BEMC and EEMC, following standard STAR jet reconstruction procedures~\cite{star:2024dlltt}.
The transverse momentum of charged tracks and the transverse energy of calorimeter towers are required to be greater than 0.2~GeV.
In addition, the tower energy $E_{\text{T}}$ in the BEMC or EEMC is corrected by subtracting the transverse momentum of a track pointed to a BEMC or EEMC tower, if its $p_{\text{T}}\cdot c$ is smaller than tower energy. If the track $p_{\text{T}}\cdot c$ is greater than the transverse energy of the tower, the tower $E_{\text{T}}$ is set to zero ~\cite{Adam_2018_LongitudinalDoublespin, _2019_LongitudinalDoublespin}. 
For anti-protons, the annihilation effects inside BEMC/EEMC material are found to be non-negligible, and the deposited energy in the $3\times3$ tower patch in the BEMC/EEMC matched to a $\overline{p}$ track is removed~\cite{star:2024dlltt}. 
This additional energy deposition can contribute to the trigger and thus influence the reconstructed jet energy distribution, which in turn leads to slight differences in the mean kinematic values of $\Lambda$ and $\overline{\Lambda}$ in a given bin.
The reconstructed jets are then corrected for the underlying-event contributions using the off-axis cone method~\cite{alice:2015ue}. 
%The jet neutral energy fraction is required to be smaller than 0.95. After underlying event correction using off-axis cone method~\cite{alice:2015ue}, 
The angular separation between $\vec{\boldsymbol{p}}_\Lambda$ and $\vec{\boldsymbol{p}}_{\text{jet}}$ is required to exceed $0.05$ to ensure sufficient resolution in determining the $\Lambda$ polarization direction.
Jets are required to have pseudorapidity relative to the event vertex in the range $|\eta_{jet}| < 1.0$ and relative to the center of STAR in the range $-0.7 < \eta_{det} < 0.9$. The reason for asymmetric $\eta_{det}$ is due to the EEMC acceptance, which only covers one side of STAR.
Jets containing at least one $\Lambda$ or $\overline{\Lambda}$ with $p_{\text{T}}^{\text{jet}} > 6$~GeV$/c$ are retained for further analysis.

The $\Lambda(\overline{\Lambda})$ polarization is extracted via the angular distribution of the daughter (anti-)proton in the $\Lambda(\overline{\Lambda})$ rest frame, following the equation:
\begin{equation}
    \frac{\text{d}N}{\mathrm{dcos}\:\theta^*} \propto \mathcal{A}(\mathrm{cos}\:\theta^*)(1 + \alpha_{\Lambda(\overline{\Lambda})}P_{\Lambda(\overline{\Lambda})}\mathrm{cos}\:\theta^*),
    \label{eq:weak_decay}
\end{equation}
where $\mathcal{A}(\mathrm{cos}\:\theta^*)$ is the detector acceptance as a function of $\cos\theta^*$, $\theta^*$ is the angle between the $\Lambda(\overline{\Lambda})$ polarization direction and its daughter $p(\overline{p})$'s momentum in the $\Lambda(\overline{\Lambda})$ rest frame, and $\alpha_{\Lambda}$ = 0.747 $\pm$ 0.009 ($\alpha_{\overline{\Lambda}}$ = $-$ 0.757 $\pm$ 0.004) is the decay parameter~\cite{PDG2024}. 
%The magnitude of $\alpha_{\overline{\Lambda}}$ is euqal to 0.757 $\pm$ 0.004 but with opposite sign.
$P_{\Lambda(\overline{\Lambda})}$ is the polarization value of interest. 
%As illustrated in figure{}~\ref{fig:Linjet}, the polarization direction is along the normal to the plane spanned by the jet axis and the $\Lambda(\overline{\Lambda})$ momentum. 

The angular distribution must be corrected for detector acceptance, $\mathcal{A}(\cos\theta^*)$, before extracting the polarization $P_{\Lambda(\overline{\Lambda})}$. To determine the acceptance, a mixed-event (ME) method is employed. 
A reconstructed $\Lambda(\overline{\Lambda})$ candidate is embedded into a different, uncorrelated event that contains a jet, reconstructed prior to the embedding, which is required to be geometrically close to the $\Lambda(\overline{\Lambda})$ candidate ($\Delta R \equiv \sqrt{(\Delta\eta)^2 + (\Delta\phi)^2} < 0.7$). Jets are then reconstructed again in the mixed event, including the embedded $\Lambda(\overline{\Lambda})$ candidate, using the same jet-finding algorithm and selection criteria as those applied to real events.
This procedure removes physical spin correlations between the $\Lambda(\overline{\Lambda})$ candidate and the jet while preserving detector acceptance effects. To minimize differences between mixed and real events, the $z$ positions of the two event vertices are required to agree within 5~cm, and both events are taken from the same run to ensure consistent beam and detector conditions. 
After these selections, residual differences between mixed and real events are observed in kinematic distributions such as the jet pseudorapidity $\eta_{\text{jet}}$ and $\Delta\eta$, $\Delta\phi$ between the jet and the $\Lambda(\overline{\Lambda})$. These differences arise because genuine correlations between the hyperon and the jet are present in real events but are intentionally absent in the ME sample. To reduce potential acceptance biases from these effects, a three-dimensional reweighting in $\eta_{\text{jet}}$, $\Delta\eta$, and $\Delta\phi$ is applied to the ME sample to match the corresponding distributions in real events. The impact of this reweighting is found to be small. 
Correlations between the jet $p_T$ and the $\Lambda(\overline{\Lambda})$ $p_T$ are mostly preserved in mixed-event reconstruction after above requirements, and the remaining small differences compared to real events have been checked and found to have negligible impact on the extracted polarization.

The ME acceptance correction is validated using a Monte Carlo sample generated with PYTHIA~6.4.28~\cite{pythia6} using the STAR-tuned Perugia~2012 parameters~\cite{perugia,_2019_LongitudinalDoublespin} and processed through the full STAR detector simulation based on GEANT 3~\cite{geant3}. A closure test is performed by introducing a known polarization signal at the generator level and extracting it at the detector level using the ME method. The reconstructed polarization agrees with the input value within uncertainties, and any residual difference is included in the systematic uncertainty.

The $\Lambda(\overline{\Lambda})$ signal region for polarization extraction is chosen as $1.112 < M < 1.120 $ GeV/$c^2$ for good purity, and the residual background is subtracted with side-band method before the acceptance corrections are applied.
The background is estimated from side-band regions [1.092, 1.102] GeV/$c^2$ and [1.130, 1.140] GeV/$c^2$, scaled according to the width of the signal period.
The same procedure for background subtraction is also performed on the ME sample.  $P_{\Lambda(\overline{\Lambda})}$ can then be extracted by fitting the acceptance-corrected $\cos\theta^*$ distribution. %with Eq. (\ref{eq:weak_decay}). 
%Only jets with minimum $p_{\text{T}}^{jet} > 6$ GeV/$c$ and $|\eta_{\text{jet}}|<1.0$ are selected, which is attributed to the detector trigger system. We require $\Delta R > $ 0.02 to reduce the contribution of a probably fake jet dominated by $\Lambda$. 
A null-test measurement is performed with the spin-0 $K_S^0$ inside jets via its decay channel $K_S^0 \rightarrow \pi^+\pi^-$. 
The same analysis procedure used for the $\Lambda(\overline{\Lambda})$ sample is applied to $K_S^0$ assuming an artificial weak decay parameter $\alpha$ = 1. The extracted ``polarizations" for $K_S^0$ are consistent with zero as expected.

The resolution in determining the $\Lambda$ polarization direction, driven by the resolution of the jet axis, is studied with the MC sample, which leads to a dilution of the polarization signal. 
In this study, detector-level jets—reconstructed with full detector acceptance effects—are matched to particle-level jets, which represent the “true” jets composed of all stable final-state particles generated by PYTHIA. The difference between the directions of the detector-level and particle-level jet axes introduces a smearing effect in determining the polarization direction. 
%This effect is particularly significant for $\Lambda$ candidates located very close to the jet axis, where small angular deviations can lead to sizable uncertainties in the polarization measurement.
%To minimize this effect, the relative distance $\Delta R$ between the $\Lambda$ and the jet is required to be greater than 0.05. 
%This selection reduces the sensitivity to jet axis resolution effects and improves the reliability of the extracted polarization.
The corresponding dilution factor is found to vary from 0.86 to 0.94 with increasing jet $p_{\text{T}}$. The measured polarization was then corrected by this dilution factor.  

%% systematic uncertainties

Several sources of systematic uncertainties are considered. The dominant uncertainty is from the jet-patch trigger, which may bias the jet flavor decomposition, as discussed in a previous publication~\cite{star:2024dlltt}. 
The trigger effects are studied using a similar MC sample as mentioned above. 
%which is generated by PYTHIA 6.4.28~\cite{pythia6} with detector responses simulated with GEANT3~\cite{geant3}. The MC events are then embedded into zero-bias events collected at STAR to account for the background environment of real data.
%To estimate potential variations in jet flavor, we simulated the effect using an embedding sample generated by PYTHIA6.4.28~\cite{pythia6}, the STAR detector simulation package based on GEANT3~\cite{geant3}, and zero-bias events recorded by the STAR detector. 
%This simulation procedure is similar to previously published STAR measurements, for example in Ref.~\cite{star:2024dlltt}.
The relative changes of quark flavor fractions between non-triggered and triggered samples are propagated to the relative changes of the polarization by conservatively assuming zero contribution from gluons.
The resulting absolute uncertainty ranges from 0.0001 to 0.0036 in different jet $p_{\text{T}}$, $z$, and $j_\text{T}$ bins. 
The second source of uncertainty arises from the ME correction to the detector acceptance. As described above, a closure test using Monte Carlo samples with an input polarization signal is performed to validate the ME method. The extracted results are consistent with the input value, and the residual absolute differences of about 0.0019 are taken as the systematic uncertainty.
Another uncertainty is due to the residual background subtraction, which is estimated by varying the side-band windows. 
%To estimate the impact of the choice of the side-band mass windows in background subtraction, we vary the side-band windows and extract the corresponding polarizations. 
The maximum change in $P_{\Lambda(\overline{\Lambda})}$ from these variations is taken as the absolute systematic uncertainty, which is at most 0.001.
The systematic uncertainty propagated from the statistical uncertainty of the dilution factor in determining the polarization direction ranges from 2.5\% to 1.0\% from the lowest to the highest jet $p_{\text{T}}$ bin. 
The last contribution is due to the uncertainty of the decay parameter $\alpha$, which is 1.2\% (0.5\%) for $\Lambda(\overline{\Lambda})$ as a relative uncertainty for all polarization results.
The above systematic uncertainties are considered to be independent and are combined in quadrature.

%physics results 

\section{Results and discussion}

\begin{figure}[tb!]
    \centering
    \includegraphics[width=0.8\linewidth]{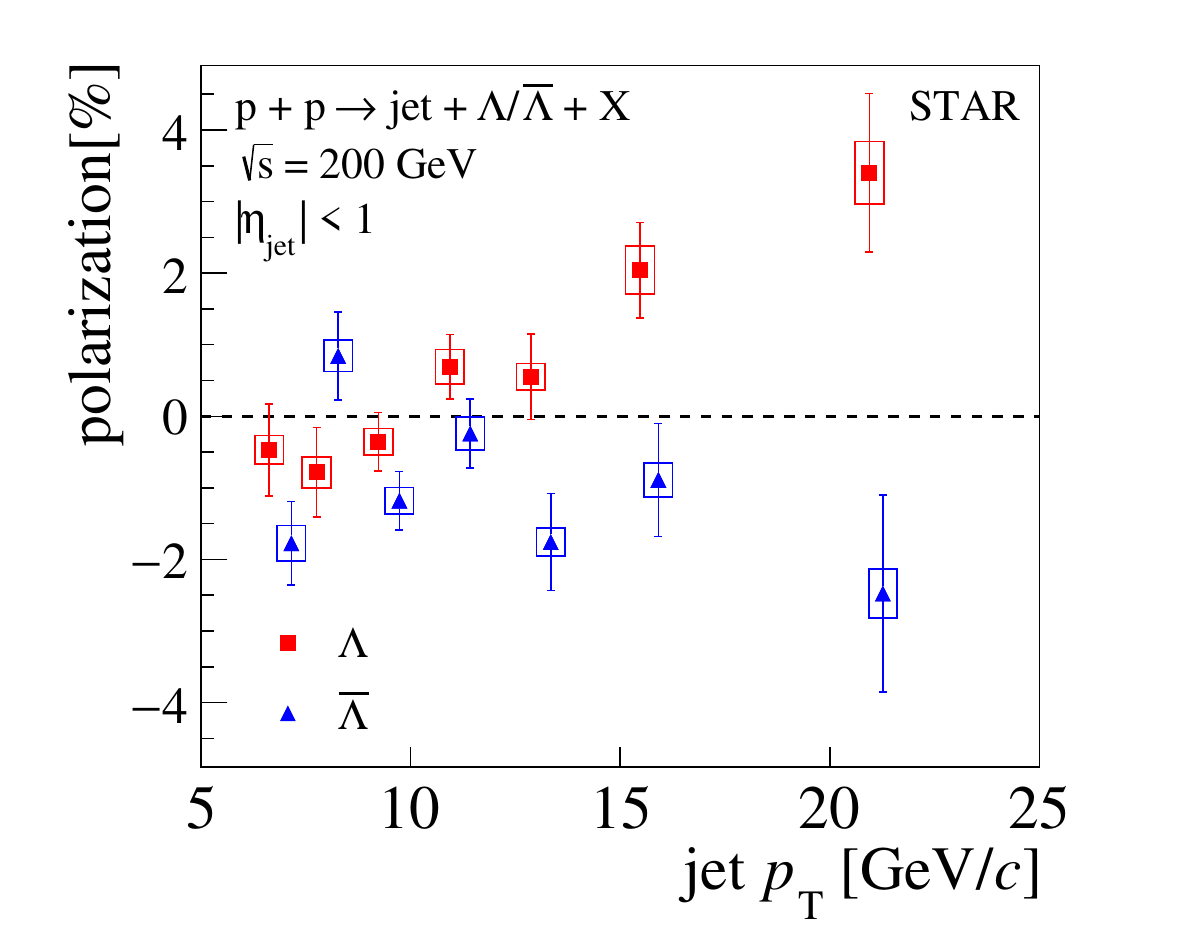}
	\caption{Transverse polarization of $\Lambda$ and $\overline{\Lambda}$ hyperons within jets as a function of particle-level jet $p_{\text{T}}$ in $pp$ collisions at $\sqrt{s}$ = 200 GeV at RHIC. The dashed line represents the baseline corresponding to zero polarization. 
    Statistical uncertainties are shown as vertical bars. Systematic uncertainties are shown as boxes. 
    \label{fig:pol_jetpt}}
\end{figure}

The transverse polarization $P$ of $\Lambda(\overline{\Lambda})$ hyperons inside jets in $pp$ collisions is first extracted as a function of the jet transverse momentum $p_{\mathrm{T}}$. The jet $p_{\mathrm{T}}$ resolution is significantly affected by detector acceptance and reconstruction effects, and therefore requires correction to enable a direct comparison with theoretical predictions. This correction is performed by matching detector-level jets to particle-level jets using  MC samples as described in previous section. Jets are matched based on their angular proximity, requiring a small $\Delta R$ between their axes. For each matched pair, the correlation between detector-level and particle-level $p_{\mathrm{T}}$ is used to construct a response matrix. The measured jet $p_{\mathrm{T}}$ distribution is then corrected using this response, following a procedure analogous to an unfolding approach adopted in Ref.~\cite{star:2024dlltt}. 
%In practice, the correction is implemented using bin-by-bin correction factors, rather than a full unfolding. 
This method accounts for both the smearing and inefficiency introduced by the detector.
%The jet $p_{\mathrm{T}}$ is corrected from the detector level to the particle level using embedded Monte Carlo samples, following the procedure adopted in Ref.~\cite{star:2024dlltt}. Here, the particle level refers to jets reconstructed with the same algorithm from all final-state particles. 
After this correction, the results can be directly compared with theoretical calculations. The values of $P_{\Lambda(\overline{\Lambda})}$ as a function of the particle-level jet $p_{\mathrm{T}}$ at $\sqrt{s}=200$~GeV within $|\eta_{\mathrm{jet}}|<1.0$ are shown in figure~\ref{fig:pol_jetpt}. In our detailed MC study from PYTHIA simulation, it is estimated that 60\% of the reconstructed $\Lambda$ hyperons originate from heavier hyperon decay. The impact from this feed-down effect is not corrected in our final results 
and should be considered for theory-data comparison.

As shown in figure~\ref{fig:pol_jetpt}, an indication of $p_T$ dependence of the $\Lambda$ polarization on jet $p_{\mathrm{T}}$ can be seen. Such a trend may arise from the varying parton-flavor composition of jets across different $p_{\mathrm{T}}$ regions, given the expected flavor dependence of the polarizing fragmentation functions~\cite{DAlesio:2020wjq,Callos:2020qtu,Chen:2021hdn}. 
In particular, the contribution from gluon fragmentation to $\Lambda(\overline{\Lambda})$ yield could be as large as 50\% in $pp$ collisions~\cite{DAlesio:2024ope}. 
Gluon contributions dominate in the low and medium jet $p_{\text{T}}$ region.
%while quark contributions increase with jet $p_{\text{T}}$~\cite{_2019_LongitudinalDoublespin}. 
The $\Lambda$ polarization shows an indication of a transition from negative values at low jet $p_{\mathrm{T}}$ to positive values at higher jet $p_{\mathrm{T}}$, with an average value of $0.24 \pm 0.19\,(\mathrm{stat}) \pm 0.09\,(\mathrm{sys})$\%. In contrast, the $\overline{\Lambda}$ polarization remains predominantly negative over the measured jet $p_{\mathrm{T}}$ range, with an average value of $-0.77 \pm 0.20\,(\mathrm{stat}) \pm 0.09\,(\mathrm{sys})$\%. PYTHIA simulation shows that the parent parton flavor composition differs between $\Lambda$ and $\overline{\Lambda}$ jets in $pp$ collisions. In particular, $\Lambda$ jets receive an increasing contribution from $u$ and $d$ quarks with jet $p_{\mathrm{T}}$, while $\overline{\Lambda}$ jets are relatively more influenced by gluons and sea antiquarks. 

The contribution of the PFF of the gluon to $P_{\Lambda(\overline{\Lambda})}$ could be significant here, but is not yet constrained by the $e^+e^-$ data ~\cite{Kang:2020xyq,DAlesio:2024ope,Gao:2024bfp}, leading to large uncertainty in the predictions for the $P_{\Lambda(\overline{\Lambda})}$ in $pp$ collision.  For example, the model prediction in Ref.~\cite{Gao:2024bfp}, based on a global fit to $e^{+}e^{-}$ data with isospin symmetry assumptions for quark fragmentation and simplified scenarios for the gluon PFF, overestimates the measured polarization by approximately an order of magnitude. The present measurements therefore provide new experimental constraints on the gluon contribution to the polarizing fragmentation function in $pp$ collisions.

% \begin{figure}[tb!]
%     \centering
%     \includegraphics[width=1.0\linewidth]{Fig/pol_vs_z.pdf}
%     \includegraphics[width=1.0\linewidth]{Fig/pol_vs_jt.pdf}
%     % \includegraphics[width=1.0\linewidth]{Fig/pol_vs_z_jt.pdf}
% 	\caption{Transverse polarization of $\Lambda$, and $\overline{\Lambda}$ as a function of $z$ and $j_{\text{T}}$ at different jet $p_{\text{T}}$ ranges [$6.2, 8.5$] GeV$/c$ (left), [$8.5, 11.9$] GeV$/c$ (middle) and $ p_{\text{T}}^{jet} > 11.9$ GeV$/c$ (right). The red and blue curves show model calculations for $\Lambda$ and $\overline{\Lambda}$ respectively from Ref.~\cite{DAlesio:2024ope}. The average polarization in each jet $p_{\text{T}}$ range is also shown with total uncertainties. 
%     \label{fig:pol_z}}
% \end{figure}

\begin{figure*}[tbh!]
    \centering
    \includegraphics[width=0.95\linewidth]{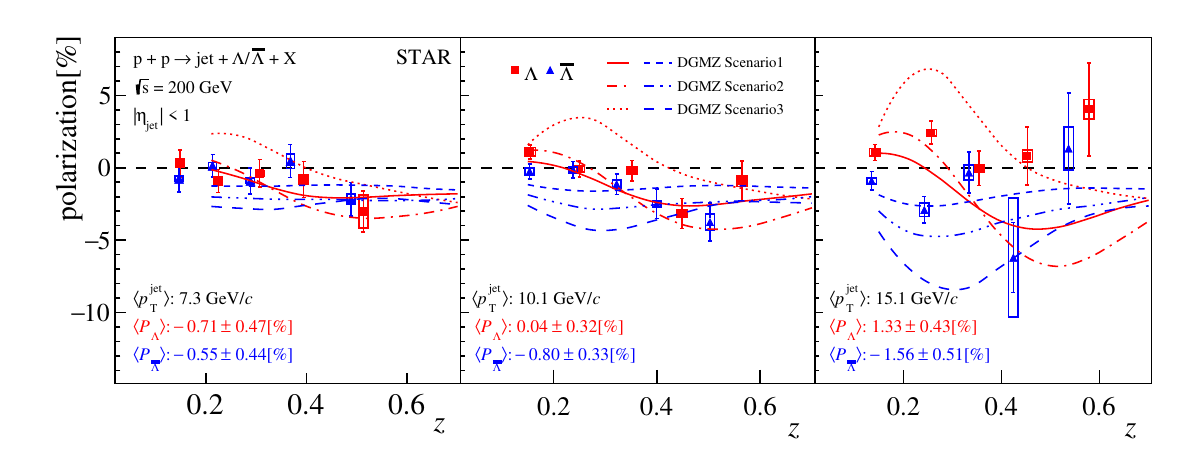}
	\caption{Transverse polarization of $\Lambda$, and $\overline{\Lambda}$ as a function of $z$ at different jet $p_{\text{T}}$ ranges [$6.2, 8.5$] GeV$/c$ (left), [$8.5, 11.9$] GeV$/c$ (middle) and $ p_{\text{T}}^{jet} > 11.9$ GeV$/c$ (right). Statistical uncertainties are shown as vertical bars. Systematic uncertainties are shown as boxes. The red and blue curves show model calculations for $\Lambda$ and $\overline{\Lambda}$ respectively from Ref.~\cite{DAlesio:2024ope}. The average polarization in each jet $p_{\text{T}}$ range is also shown with total uncertainties. 
    \label{fig:pol_z}}
\end{figure*}

\begin{figure*}[bth!]
    \centering
    \includegraphics[width=0.95\linewidth]{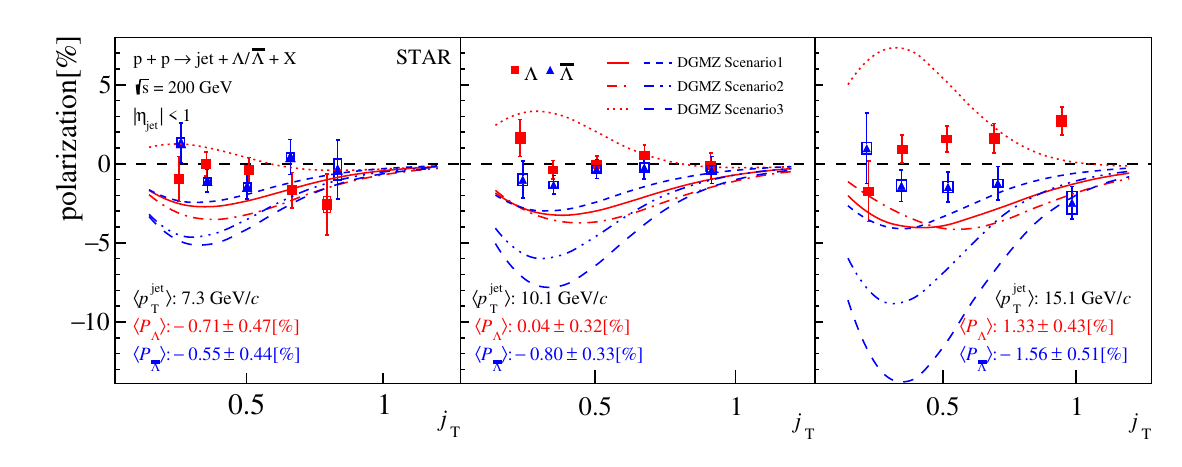}
	\caption{Transverse polarization of $\Lambda$ and $\overline{\Lambda}$ hyperons as a function of $j_{\text{T}}$ for different jet $p_{\text{T}}$ ranges of [$6.2, 8.5$] GeV$/c$ (left), [$8.5, 11.9$] GeV$/c$ (middle) and $ p_{\text{T}}^{jet} > 11.9$ GeV$/c$ (right). 
    %Here $ p_{\text{T}}^{jet}$ binning is based on detector level, and the average $p_{\text{T}}$ in each range is corrected to particle level. 
    Statistical uncertainties are shown as vertical bars. Systematic uncertainties are shown as boxes.  The red and blue curves show model calculations for $\Lambda$ and $\overline{\Lambda}$ respectively from Ref.~\cite{DAlesio:2024ope}. \label{fig:pol_jt}}
\end{figure*}

To provide more constraints on the PFF, both collinear and transverse momentum dependent, the transverse polarizations of $\Lambda$ and $\overline{\Lambda}$ are also measured as functions of $z$ and $j_{\text{T}}$, as shown in figure{}~\ref{fig:pol_z} and~\ref{fig:pol_jt}, respectively. 
Both $z$ and $j_{\text{T}}$ are corrected from detector level to particle level, as is done for jet $p_{\text{T}}$.
Since the $\Lambda$ polarization changes sign from negative to positive with increasing jet $p_{\text{T}}$, the average value of polarization, $z$ and $j_{\text{T}}$ dependencies are shown in three jet $p_{\text{T}}$ ranges. 
%Therefore, we divided jet $p_{\text{T}}$ into three ranges: $6 < p_{\text{T}}^{jet} < 8.4$ GeV, $8.4 < p_{\text{T}}^{jet} < 12$ GeV, and $p_{\text{T}}^{jet} > 12$ GeV. 
%The polarizations of $\Lambda$ and $\overline{\Lambda}$ exhibit different $z$ dependencies across the jet $p_{\text{T}}$ ranges. 
In the low jet $p_{\text{T}}$ range $6.2 < p_{\text{T}}^{jet} < 8.5$ GeV$/c$ and medium $p_{\text{T}}$ range $8.5 < p_{\text{T}}^{jet} < 11.9$ GeV$/c$, neither $\Lambda$ nor $\overline{\Lambda}$ polarization shows a clear dependence on $z$.  
In the highest jet $p_{\text{T}}$ range $p_{\text{T}}^{jet} > 11.9$ GeV$/c$, the polarizations of $\Lambda$ and $\overline{\Lambda}$ are mostly opposite in sign,
as can also be seen from the average value shown in the figure. 
On the $j_{\text{T}}$ dependence in the lower panel, no significant dependence is observed for $\Lambda$ and $\overline{\Lambda}$ in all $p_{\text{T}}$ ranges, rather than the clear opposite sign for $\Lambda$ and $\overline{\Lambda}$ in the highest $p_{\text{T}}$ range. 
Both $u$, $d$ quarks fragmentation contribute significantly to $\Lambda$ and $\overline\Lambda$ in $pp$, but in a very different way for PFF contributions, because $u$, $d$ are valence quarks for $\Lambda$ while sea quarks for $\overline\Lambda$. 

%In Figs.~\ref{fig:pol_z} and \ref{fig:pol_jt}, the data are compared with theoretical predictions~\cite{DAlesio:2024ope} using three different parameterizations of the PFFs (DGMZ scenarios 1–3) based on BELLE data~\cite{Belle:2018ttu}, with the gluon PFF set to zero. 
In figure~\ref{fig:pol_z} and figure~\ref{fig:pol_jt}, the data are compared with theoretical predictions~\cite{DAlesio:2024ope} using three different parameterizations of the PFFs (DGMZ scenarios 1–3) based on BELLE data~\cite{Belle:2018ttu}, with the gluon PFF set to zero. 
DGMZ Scenario 1 includes $u, d, s$ and their antiquark contributions without SU(2) isospin asymmetry; Scenario 2 adds charm contribution based on Scenario 1; Scenario 3 includes SU(2) isospin asymmetry~\cite{DAlesio:2024ope}. 
As seen from figure~\ref{fig:pol_z}, there is reasonable agreement between data and model predictions versus $z$ in the lowest jet $p_{\text{T}}$ range. Sizable discrepancies can be seen in the highest jet $p_{\text{T}}$ range versus $z$. 
Also, significant differences between data and model predictions are seen for the results versus $j_{\text{T}}$ in particular in the higher jet $p_{\text{T}}$ range.  
Another theoretical prediction in Ref.~\cite{Kang:2020xyq} on $\Lambda$ polarization in jets in $pp$ also uses only quark PFFs constrained by BELLE data~\cite{Belle:2018ttu}, giving mostly negative $P_\Lambda$ versus $z$ at a different kinematic range ($10 < p_{\text{T}}^{jet} < 15$ GeV/$c$), with similar magnitude to our $\Lambda$ data.
Again, the knowledge of gluon PFF is crucial in understanding the $pp$ data. 
% in the middle panel of figure~\ref{fig:pol_z}, $i.e.$, when $10 < p_{\text{T}}^{jet} < 15$ GeV/$c$.  

%the theoretical curves describing $e^+e^-$ data fail to describe $pp$ results, underscoring the necessity of $pp$ collision data for constraining PFFs.
%: scenarios 1 only include $u\ d\ s$ and their antiquark contributions; scenarios 2 add charm conrtibution based on scenarios 1; scenarios 3 inclusive SU(2). 
%Apparently, the theoretical curves fitted  $e^+e^-$ data can not describe $pp$ data, which confirms our results in $pp$ collisions is necessary for constraints of PFF.

The PFFs are important in understanding the large transverse polarizations observed relative to the production plane in hadron-hadron collisions, as mentioned in the introduction. 
The polarization relative to the production plane is expected to be small at mid-rapidity for low $x_F$ ($=2p_z/\sqrt s$) in hadron-hadron collisions~\cite{Bunce:1976yb,Panagiotou:1989sv}, but the polarization inside a jet could still be non-zero~\cite{Boer:2007nh}, as confirmed for the first time by our measurements.

%%% Summary 
\section{Conclusions}
In summary, we report the first measurements of the transverse polarization of $\Lambda(\overline{\Lambda})$ inside jets in unpolarized $pp$ collisions at $\sqrt{s} = 200$ GeV. 
The measurements directly probe the polarizing fragmentation function, which is an important contribution to the surprisingly large transverse $\Lambda$ polarizations known for 50 years in hadron-hadron collisions.
 %indicating that the spontaneous polarization of $\Lambda$ may have a flavor dependence. 
An indication of jet $p_{\text{T}}$ dependence in the transverse polarization of $\Lambda$ is observed.
The $\Lambda(\overline{\Lambda})$ polarizations are also measured as functions of $z$ and $j_{\text{T}}$ in different jet $p_{\text{T}}$ ranges, providing constraints for the polarizing TMD fragmentation functions. 
%The dependencies of $\Lambda(\overline{\Lambda})$ polarization versus longitudinal momentum fraction $z$ and transverse momentum $j_T$ are also measured in different jet $p_{\text{T}}$ ranges. 
%
These results will provide the first constraints on the gluon PFF, which is not yet constrained by $e^+e^-$ annihilation data. 
The sizable discrepancy between data and model predictions reveals the importance of $pp$ data.
The reported data furthermore cover a wide range of jet energy.
Taken together, these measurements will provide crucial constraints on polarizing fragmentation functions including TMD evolution effects, and test the universality in different processes when combined with data from $e^+e^-$ and DIS facilities, including the future Electron-Ion Collider~\cite{AbdulKhalek:2021gbh}. 

%%Acknowledgement
\acknowledgments
We would like to thank U. D'Alesio, M. Zaccheddu and Yukun Song for providing the results of their theoretical calculations. We thank the RHIC Operations Group and SCDF at BNL, the NERSC Center at LBNL, and the Open Science Grid consortium for providing resources and support.  This work was supported in part by the Office of Nuclear Physics within the U.S. DOE Office of Science, the U.S. National Science Foundation, National Natural Science Foundation of China, Chinese Academy of Science, the Ministry of Science and Technology of China and the Chinese Ministry of Education, NSTC Taipei, the National Research Foundation of Korea, Czech Science Foundation and Ministry of Education, Youth and Sports of the Czech Republic, Hungarian National Research, Development and Innovation Office, New National Excellency Programme of the Hungarian Ministry of Human Capacities, Department of Atomic Energy and Department of Science and Technology of the Government of India, the National Science Centre and WUT ID-UB of Poland, German Bundesministerium f\"ur Bildung, Wissenschaft, Forschung and Technologie (BMBF), Helmholtz Association, Ministry of Education, Culture, Sports, Science, and Technology (MEXT), Japan Society for the Promotion of Science (JSPS), and Agencia Nacional de Investigacion y Desarrollo de Chile (ANID), Chile.  
\bibliographystyle{JHEP}
\bibliography{biblio.bib}

@article{Bunce:1976yb,
    author = "Bunce, G. and others",
    title = "{$\Lambda^0$ Hyperon Polarization in Inclusive Production by 300-{GeV} Protons on Beryllium.}",
    reportNumber = "FERMILAB-PUB-76-157-E, D76-04710",
    doi = "10.1103/PhysRevLett.36.1113",
    journal = "Phys. Rev. Lett.",
    volume = "36",
    pages = "1113--1116",
    year = "1976"
}

@article{pQCD_cal,
  title = {Transverse Quark Polarization in Large-${p}_{T}$ Reactions, ${e}^{+}{e}^{-}$ Jets, and Leptoproduction: A Test of Quantum Chromodynamics},
  author = {Kane, G. L. and Pumplin, J. and Repko, W.},
  journal = {Phys. Rev. Lett.},
  volume = {41},
  issue = {25},
  pages = {1689--1692},
  numpages = {0},
  year = {1978},
  month = {Dec},
  publisher = {American Physical Society},
  doi = {10.1103/PhysRevLett.41.1689},
  url = {https://link.aps.org/doi/10.1103/PhysRevLett.41.1689}
}

@article{Panagiotou:1989sv,
    author = "Panagiotou, Apostolos D.",
    title = "{$\Lambda^0$ Polarization in Hadron - Nucleon, Hadron - Nucleus and Nucleus-nucleus Interactions}",
    reportNumber = "CERN-EP-89-131",
    doi = "10.1142/S0217751X90000568",
    journal = "Int. J. Mod. Phys. A",
    volume = "5",
    pages = "1197",
    year = "1990"
}

@article{Felix:1999tf,
    author = "Felix, J.",
    title = "{On Theoretical studies of $\Lambda^0$ polarization}",
    doi = "10.1142/S0217732399000870",
    journal = "Mod. Phys. Lett. A",
    volume = "14",
    pages = "827--842",
    year = "1999"
}

@article{NOMAD:2000wdf,
    author = "Astier, P. and others",
    collaboration = "NOMAD",
    title = "{Measurement of the Lambda polarization in nu/mu charged current interactions in the NOMAD experiment}",
    reportNumber = "CERN-EP-2000-111",
    doi = "10.1016/S0550-3213(00)00503-4",
    journal = "Nucl. Phys. B",
    volume = "588",
    pages = "3--36",
    year = "2000"
}

@article{NOMAD:2001iup,
    author = "Astier, P. and others",
    collaboration = "NOMAD",
    title = "{Measurement of the anti-Lambda polarization in muon-neutrino charged current interactions in the NOMAD experiment}",
    eprint = "hep-ex/0103047",
    archivePrefix = "arXiv",
    reportNumber = "CERN-EP-2001-028",
    doi = "10.1016/S0550-3213(01)00181-X",
    journal = "Nucl. Phys. B",
    volume = "605",
    pages = "3--14",
    year = "2001"
}

@article{HERMES:2007fpi,
    author = "Airapetian, A. and others",
    collaboration = "HERMES",
    title = "{Transverse Polarization of Lambda and anti-Lambda Hyperons in Quasireal Photoproduction}",
    eprint = "0704.3133",
    archivePrefix = "arXiv",
    primaryClass = "hep-ex",
    reportNumber = "DESY-07-036",
    doi = "10.1103/PhysRevD.76.092008",
    journal = "Phys. Rev. D",
    volume = "76",
    pages = "092008",
    year = "2007"
}

@article{ATLAS:2014ona,
    author = "Aad, Georges and others",
    collaboration = "ATLAS",
    title = "{Measurement of the transverse polarization of $\Lambda$ and $\overline{\Lambda}$ hyperons produced in proton-proton collisions at $\sqrt{s}=7$ TeV using the ATLAS detector}",
    reportNumber = "CERN-PH-EP-2014-258",
    doi = "10.1103/PhysRevD.91.032004",
    journal = "Phys. Rev. D",
    volume = "91",
    number = "3",
    pages = "032004",
    year = "2015"
}

@article{LHCb:2024vwi,
    author = "Aaij, R. and others",
    collaboration = "LHCb",
    title = "{Transverse polarization measurement of $\Lambda$ hyperons in pNe collisions at $ \sqrt{s_{NN}} $ = 68.4 {GeV} with the LHCb detector}",
    reportNumber = "CERN-EP-2024-121, LHCb-PAPER-2024-009",
    doi = "10.1007/JHEP09(2024)082",
    journal = "JHEP",
    volume = "09",
    pages = "082",
    year = "2024"
}

@misc{LHCb:2025Lambda,
    title="{Measurement of transverse $\Lambda$ and $\overline{\Lambda}$ hyperon polarization in $p$Pb collisions at $\sqrt{s_{NN}} = 5.02$ TeV}", 
    author = "Aaij, R. and others",    
    collaboration = "LHCb",
      year={2025},
      eprint={2508.02009},
      archivePrefix={arXiv},
      primaryClass={nucl-ex},
}

@article{STAR:2025njp,
    author = "Aboona, B. E. and others",
    collaboration = "STAR",
    title = "{Measuring spin correlation between quarks during QCD confinement}",
    eprint = "2506.05499",
    archivePrefix = "arXiv",
    primaryClass = "hep-ex",
    doi = "10.1038/s41586-025-09920-0",
    journal = "Nature",
    volume = "650",
    pages = "65--71",
    year = "2026"
}

@article{Chen:2026gka,
	author = "Chen, Jinhui and others",
	doi = {10.1088/0256-307X/43/3/030102},
	journal = {Chin. Phys. Lett.},
	number = {3},
	pages = {030102},
	title = {Selected highlights from {STAR} experiment},
	url = {http://cpl.iphy.ac.cn/en/article/doi/10.1088/0256-307X/43/3/030102},
	volume = {43},
	year = {2026}}

@article{Boer:1997nt,
    author = "Boer, Daniel and Mulders, P. J.",
    title = "{Time reversal odd distribution functions in leptoproduction}",
    reportNumber = "NIKHEF-97-049, VUTH-97-20",
    doi = "10.1103/PhysRevD.57.5780",
    journal = "Phys. Rev. D",
    volume = "57",
    pages = "5780--5786",
    year = "1998"
}

@article{Liang:1997rt,
    author = "Liang, Zuo-Tang and Boros, C.",
    title = "{Hyperon polarization and single spin left-right asymmetry in inclusive production processes at high-energies}",
    doi = "10.1103/PhysRevLett.79.3608",
    journal = "Phys. Rev. Lett.",
    volume = "79",
    pages = "3608--3611",
    year = "1997"
}

@article{Koike:2017fxr,
    author = "Koike, Yuji and Metz, Andreas and Pitonyak, Daniel and Yabe, Kenta and Yoshida, Shinsuke",
    title = "{Twist-3 fragmentation contribution to polarized hyperon production in unpolarized hadronic collisions}",
    doi = "10.1103/PhysRevD.95.114013",
    journal = "Phys. Rev. D",
    volume = "95",
    number = "11",
    pages = "114013",
    year = "2017"
}

@article{Koike:2022ddx,
    author = "Koike, Yuji and Takada, Kazuki and Usui, Sumire and Yabe, Kenta and Yoshida, Shinsuke",
    title = "{Transverse polarization of hyperons produced in semi-inclusive deep inelastic scattering}",
    doi = "10.1103/PhysRevD.105.056021",
    journal = "Phys. Rev. D",
    volume = "105",
    number = "5",
    pages = "056021",
    year = "2022"
}

@article{Ikarashi:2022zeo,
    author = "Ikarashi, Riku and Koike, Yuji and Yabe, Kenta and Yoshida, Shinsuke",
    title = "{New derivation of the twist-3 gluon fragmentation contribution to polarized hyperon production}",
    doi = "10.1103/PhysRevD.106.074006",
    journal = "Phys. Rev. D",
    volume = "106",
    number = "7",
    pages = "074006",
    year = "2022"
}

@article{Kanazawa:2000cx,
    author = "Kanazawa, Y. and Koike, Yuji",
    title = "{Polarization in hadronic Lambda hyperon production and chiral odd twist - three quark distribution}",
    doi = "10.1103/PhysRevD.64.034019",
    journal = "Phys. Rev. D",
    volume = "64",
    pages = "034019",
    year = "2001"
}

@article{Anselmino:2000vs,
    author = "Anselmino, M. and Boer, Daniel and D'Alesio, U. and Murgia, F.",
    title = "{Lambda polarization from unpolarized quark fragmentation}",
    reportNumber = "DFTT-32-2000, INFN-CA-TH-0012",
    doi = "10.1103/PhysRevD.63.054029",
    journal = "Phys. Rev. D",
    volume = "63",
    pages = "054029",
    year = "2001"
}

@article{Anselmino:2001js,
    author = "Anselmino, M. and Boer, Daniel and D'Alesio, U. and Murgia, F.",
    title = "{Transverse lambda polarization in semiinclusive DIS}",
    reportNumber = "DFTT-28-2001, INFNCA-TH0108",
    doi = "10.1103/PhysRevD.65.114014",
    journal = "Phys. Rev. D",
    volume = "65",
    pages = "114014",
    year = "2002"
}

@article{Zhou:2008fb,
    author = "Zhou, Jian and Yuan, Feng and Liang, Zuo-Tang",
    title = "{Hyperon Polarization in Unpolarized Scattering Processes}",
    doi = "10.1103/PhysRevD.78.114008",
    journal = "Phys. Rev. D",
    volume = "78",
    pages = "114008",
    year = "2008"
}

@article{Boer:2007nh,
    author = "Boer, D. and Bomhof, C. J. and Hwang, D. S. and Mulders, P. J.",
    title = "{Spin asymmetries in jet-hyperon production at LHC}",
    doi = "10.1016/j.physletb.2007.10.059",
    journal = "Phys. Lett. B",
    volume = "659",
    pages = "127--136",
    year = "2008"
}

@article{Mulders:1995dh,
    author = "Mulders, P. J. and Tangerman, R. D.",
    title = "{The Complete tree level result up to order $1/Q$ for polarized deep inelastic leptoproduction}",
    reportNumber = "NIKHEF-95-053",
    doi = "10.1016/0550-3213(95)00632-X",
    journal = "Nucl. Phys. B",
    volume = "461",
    pages = "197--237",
    year = "1996",
    note = "[Erratum: Nucl.Phys.B 484, 538--540 (1997)]"
}

@article{1998OPAL,
   title={Polarization and forward-backward asymmetry of {$\Lambda$} baryons in hadronic {$Z^0$} decays},
   volume={2},
   ISSN={1434-6052},
   url={http://dx.doi.org/10.1007/s100520050123},
   DOI={10.1007/s100520050123},
   number={1},
   journal={Eur. Phys. J. C},
   publisher={Springer Science and Business Media LLC},
   author={Ackerstaff et al., K.},
   year={1998},
   month=mar, pages={49–59}
}

@article{ALEPH:1996oew,
    author = "Buskulic, D. and others",
    collaboration = "ALEPH",
    title = "{Measurement of Lambda polarization from $Z$ decays}",
    reportNumber = "CERN-PPE-96-004, CERN-PPE-96-4, CERN-PPE-96-04",
    doi = "10.1016/0370-2693(96)00300-0",
    journal = "Phys. Lett. B",
    volume = "374",
    pages = "319--330",
    year = "1996"
}

@article{DAlesio:2020wjq,
    author = "D'Alesio, Umberto and Murgia, Francesco and Zaccheddu, Marco",
    title = "{First extraction of the $\Lambda$ polarizing fragmentation function from Belle $e^+e^-$ data}",
    primaryClass = "hep-ph",
    doi = "10.1103/PhysRevD.102.054001",
    journal = "Phys. Rev. D",
    volume = "102",
    number = "5",
    pages = "054001",
    year = "2020"
}

@article{Callos:2020qtu,
    author = "Callos, Daniel and Kang, Zhong-Bo and Terry, John",
    title = "{Extracting the transverse momentum dependent polarizing fragmentation functions}",
    primaryClass = "hep-ph",
    doi = "10.1103/PhysRevD.102.096007",
    journal = "Phys. Rev. D",
    volume = "102",
    number = "9",
    pages = "096007",
    year = "2020"
}

@article{Chen:2021hdn,
    author = "Chen, Kai-Bao and Liang, Zuo-Tang and Pan, Yan-Lei and Song, Yu-Kun and Wei, Shu-Yi",
    title = "{Isospin Symmetry of Fragmentation Functions}",
    primaryClass = "hep-ph",
    doi = "10.1016/j.physletb.2021.136217",
    journal = "Phys. Lett. B",
    volume = "816",
    pages = "136217",
    year = "2021"
}

@article{Li:2020oto,
    author = "Li, Hui and Wang, Xiaoyu and Yang, Yongliang and Lu, Zhun",
    title = "{The transverse polarization of $\Lambda $ hyperons in $e^+e^-\rightarrow \Lambda ^\uparrow h X$ processes within TMD factorization}",
    doi = "10.1140/epjc/s10052-021-09064-1",
    journal = "Eur. Phys. J. C",
    volume = "81",
    number = "4",
    pages = "289",
    year = "2021"
}

@article{Gamberg:2021iat,
    author = "Gamberg, Leonard and Kang, Zhong-Bo and Shao, Ding Yu and Terry, John and Zhao, Fanyi",
    title = "{Transverse $\Lambda$ polarization in $e^{+} e^{-}$ collisions}",
    doi = "10.1016/j.physletb.2021.136371",
    journal = "Phys. Lett. B",
    volume = "818",
    pages = "136371",
    year = "2021"
}

@article{Belle:2018ttu,
    author = "Guan, Y. and others",
    collaboration = "Belle",
    title = "{Observation of Transverse $\Lambda/\overline{\Lambda}$ Hyperon Polarization in $e^+e^-$ Annihilation at Belle}",
    reportNumber = "Belle Preprint 2018-17; KEK Preprint 2018-23",
    doi = "10.1103/PhysRevLett.122.042001",
    journal = "Phys. Rev. Lett.",
    volume = "122",
    number = "4",
    pages = "042001",
    year = "2019"
}

@article{Kang:2020xyq,
    author = "Kang, Zhong-Bo and Lee, Kyle and Zhao, Fanyi",
    title = "{Polarized jet fragmentation functions}",
    doi = "10.1016/j.physletb.2020.135756",
    journal = "Phys. Lett. B",
    volume = "809",
    pages = "135756",
    year = "2020"
}

@article{DAlesio:2022brl,
    author = "D'Alesio, Umberto and Gamberg, Leonard and Murgia, Francesco and Zaccheddu, Marco",
    title = "{Transverse {$\Lambda$} polarization in $e^{+}e^{-}$ processes within a TMD factorization approach and the polarizing fragmentation function}",
    doi = "10.1007/JHEP12(2022)074",
    journal = "JHEP",
    volume = "12",
    pages = "074",
    year = "2022"
}

@article{DAlesio:2023ozw,
    author = "D'Alesio, Umberto and Gamberg, Leonard and Murgia, Francesco and Zaccheddu, Marco",
    title = "{Transverse {$\Lambda$} polarization in $e^+e^-$ annihilations and in SIDIS processes at the EIC within TMD factorization}",
    doi = "10.1103/PhysRevD.108.094004",
    journal = "Phys. Rev. D",
    volume = "108",
    number = "9",
    pages = "094004",
    year = "2023"
}

@article{Kang:2021kpt,
    author = "Kang, Zhong-Bo and Terry, John and Vossen, Anselm and Xu, Qing-Hua and Zhang, Jinlong",
    title = "{Transverse Lambda production at the future Electron-Ion Collider}",
    doi = "10.1103/PhysRevD.105.094033",
    journal = "Phys. Rev. D",
    volume = "105",
    number = "9",
    pages = "094033",
    year = "2022"
}

@article{Chen:2021zrr,
    author = "Chen, Kai-Bao and Liang, Zuo-Tang and Song, Yu-Kun and Wei, Shu-Yi",
    title = "{Longitudinal and transverse polarizations of ${\Lambda}$ hyperon in unpolarized SIDIS and $e^+e^-$ annihilation}",
    doi = "10.1103/PhysRevD.105.034027",
    journal = "Phys. Rev. D",
    volume = "105",
    number = "3",
    pages = "034027",
    year = "2022"
}

@article{Gao:2024dxl,
    author = "Gao, Taoya",
    collaboration = "STAR",
    title = "{Measurement of transverse polarization of $\Lambda/\overline{\Lambda}$ within jet in $pp$ collisions at STAR}",
    doi = "10.22323/1.456.0031",
    journal = "PoS",
    volume = "SPIN2023",
    pages = "031",
    year = "2024"
}

@article{Kang:2017glf,
    author = "Kang, Zhong-Bo and Liu, Xiaohui and Ringer, Felix and Xing, Hongxi",
    title = "{The transverse momentum distribution of hadrons within jets}",
    doi = "10.1007/JHEP11(2017)068",
    journal = "JHEP",
    volume = "11",
    pages = "068",
    year = "2017"
}

@article{Metz:2002iz,
    author = "Metz, A.",
    title = "{Gluon-exchange in spin-dependent fragmentation}",
    doi = "10.1016/S0370-2693(02)02899-X",
    journal = "Phys. Lett. B",
    volume = "549",
    pages = "139--145",
    year = "2002"
}

@article{Collins:2004nx,
    author = "Collins, John C. and Metz, Andreas",
    title = "{Universality of soft and collinear factors in hard-scattering factorization}",
    doi = "10.1103/PhysRevLett.93.252001",
    journal = "Phys. Rev. Lett.",
    volume = "93",
    pages = "252001",
    year = "2004"
}

@article{Meissner:2008yf,
    author = "Meissner, S. and Metz, A.",
    title = "{Partonic pole matrix elements for fragmentation}",
    doi = "10.1103/PhysRevLett.102.172003",
    journal = "Phys. Rev. Lett.",
    volume = "102",
    pages = "172003",
    year = "2009"
}

@article{Boer:2010ya,
    author = "Boer, Daniel and Kang, Zhong-Bo and Vogelsang, Werner and Yuan, Feng",
    title = "{Test of the Universality of Naive-time-reversal-odd Fragmentation Functions}",
    doi = "10.1103/PhysRevLett.105.202001",
    journal = "Phys. Rev. Lett.",
    volume = "105",
    pages = "202001",
    year = "2010"
}

@article{STAR_det,
	author = {Ackermann, K. H. and others},
    title = {{STAR Detector Overview}},
    journal = {{Nucl. Instrum. Methods Phys. Res., Sect. A}},
    pages = {624--632},
    volume = {$\bf{499}$},
	year = {2003},
	date = {2003-03},
	doi = {10.1016/S0168-9002(02)01960-5},
	issn = {01689002}
}

@article{BEMC,
	author = {Beddo, M. and others},
	year = {2003},
	date = {2003-03},
	doi = {10.1016/S0168-9002(02)01970-8},
	issn = {01689002},
	
        journal = {{Nucl. Instrum. Methods Phys. Res., Sect. A}},
	langid = {english},
	number = {2-3},
	options = {useprefix=true},
	pages = {725--739},
	title = {{The STAR Barrel Electromagnetic Calorimeter}},
	url = {https://linkinghub.elsevier.com/retrieve/pii/S0168900202019708},
	urldate = {2020-11-25},
	volume = {$\bf{499}$}}

@article{EEMC,
	author = {Allgower, C. E. and others},
	year = {2003},
	date = {2003-03},
	doi = {10.1016/S0168-9002(02)01971-X},
	issn = {01689002},
    journal = {{Nucl. Instrum. Methods Phys. Res., Sect. A}},
	langid = {english},
	number = {2-3},
	pages = {740--750},
	title = {{The STAR Endcap Electromagnetic Calorimeter}},
	url = {https://linkinghub.elsevier.com/retrieve/pii/S016890020201971X},
	urldate = {2020-11-25},
	volume = {$\bf{499}$}}

@article{TPC,
	author = {Anderson, M. and others},
	year = {2003},
	date = {2003-03},
	doi = {10.1016/S0168-9002(02)01964-2},
	issn = {01689002},	
    journal = {{Nucl. Instrum. Methods Phys. Res., Sect. A}},
	langid = {english},
	number = {2-3},
	pages = {659--678},
	shorttitle = {The {{STAR}} Time Projection Chamber},
	title = {{The STAR Time Projection Chamber: A Unique Tool for Studying High Multiplicity Events at RHIC}},
	url = {https://linkinghub.elsevier.com/retrieve/pii/S0168900202019642},
	urldate = {2020-12-25},
	volume = {$\bf{499}$}
}

@article{star:2012jet,
  title = {Longitudinal and transverse spin asymmetries for inclusive jet production at mid-rapidity in polarized $p+p$ collisions at $\sqrt{s}=200$ {GeV}},
  author = {Adamczyk, L. and others},
  collaboration = {STAR},
  journal = {Phys. Rev. D},
  volume = {86},
  issue = {3},
  pages = {032006},
  numpages = {18},
  year = {2012},
  month = {Aug},
  publisher = {American Physical Society},
  doi = {10.1103/PhysRevD.86.032006},
  url = {https://link.aps.org/doi/10.1103/PhysRevD.86.032006}
}

@article{star:2018dll,
  title = {{Improved measurement of the longitudinal spin transfer to $\Lambda$ and $\overline{\Lambda}$ hyperons in polarized proton-proton collisions at $\sqrt{s}=200$ GeV}},
  author = {Adam, J. and others},
  collaboration = {STAR},
  journal = {Phys. Rev. D},
  volume = {98},
  issue = {11},
  pages = {112009},
  numpages = {8},
  year = {2018},
  month = {Dec},
  publisher = {American Physical Society},
  doi = {10.1103/PhysRevD.98.112009},
  url = {https://link.aps.org/doi/10.1103/PhysRevD.98.112009}
}

@article{star:2018dtt,
  title = {{Transverse spin transfer to $\Lambda$ and $\overline{\Lambda}$ hyperons in polarized proton-proton collisions at $\sqrt{s}=200$ GeV}},
  author = {Adam, J. and others},
  collaboration = {STAR},
  journal = {Phys. Rev. D},
  volume = {98},
  issue = {9},
  pages = {091103},
  numpages = {8},
  year = {2018},
  month = {Nov},
  publisher = {American Physical Society},
  doi = {10.1103/PhysRevD.98.091103},
  url = {https://link.aps.org/doi/10.1103/PhysRevD.98.091103}
}

@article{star:2024dlltt,
    author = "Abdulhamid, Muhammad and others",
    collaboration = "STAR",
    title = {{Longitudinal and transverse spin transfer to $\Lambda$ and $\overline{\Lambda}$ hyperons in polarized p+p collisions at $\sqrt{s}=200$ GeV}},
    doi = "10.1103/PhysRevD.109.012004",
    journal = "Phys. Rev. D",
    volume = "109",
    number = "1",
    pages = "012004",
    year = "2024"
}

@article{star:2009dll,
  title = {Longitudinal spin transfer to {$\Lambda$} and $\overline{\Lambda}$ hyperons in polarized proton-proton collisions at $\sqrt{s}=200$ {GeV}},
  author = {Abelev, B. I. and others},
  collaboration = {STAR},
  journal = {Phys. Rev. D},
  volume = {80},
  issue = {11},
  pages = {111102},
  numpages = {7},
  year = {2009},
  month = {Dec},
  publisher = {American Physical Society},
  doi = {10.1103/PhysRevD.80.111102},
  url = {https://link.aps.org/doi/10.1103/PhysRevD.80.111102}
}

@article{antiKt:2008gp,
	author = "Cacciari, Matteo and Salam, Gavin P. and Soyez, Gregory",
    title = "{The anti-$k_t$ jet clustering algorithm}",
    reportNumber = "LPTHE-07-03",
    doi = "10.1088/1126-6708/2008/04/063",
    journal = "JHEP",
    volume = "04",
    pages = "063",
    year = {2008}
}

@article{Adam_2018_LongitudinalDoublespin,
  title = {{Longitudinal Double-Spin Asymmetries for Dijet production at Intermediate Pseudorapidity in Polarized $pp$ Collisions at $\sqrt{s} = 200$ GeV}},
  author = {Adam, J. and others},
  collaboration = "STAR",
  date = {2018-08-17},
  year = {2018},
  journal = {Phys. Rev. D},
  volume = {$\bf{98}$},
  number = {3},
  pages = {032011},
  issn = {2470-0010, 2470-0029},
  doi = {10.1103/PhysRevD.98.032011},
  url = {https://link.aps.org/doi/10.1103/PhysRevD.98.032011},
  urldate = {2020-05-27},
  langid = {english}
}

@article{_2019_LongitudinalDoublespin,
  author = {Adam, J. and others},
    collaboration = "STAR",
  title = {{Longitudinal Double-Spin Asymmetry for Inclusive Jet and Dijet Production in $pp$ Collisions at $\sqrt{s}=510$ GeV}},
  date = {2019},
  year = {2019},
  journal = {Phys. Rev. D},
  volume = {$\bf{100}$},
  pages = {052005},
  langid = {english}
}

@article{PDG2024,
  title = {Review of Particle Physics},
  author = {Zyla, P. A. and others},
  collaboration = {Particle Data Group},
  journal = {Phys. Rev. D},
  volume = {110},
  issue = {3},
  pages = {030001},
  numpages = {5},
  year = {2024},
  month = {Aug},
  publisher = {American Physical Society},
  doi = {10.1103/PhysRevD.110.030001},
  url = {https://link.aps.org/doi/10.1103/PhysRevD.110.030001}
}

@article{alice:2015ue,
  title = {Charged jet cross sections and properties in proton-proton collisions at $\sqrt{s}=7$ {TeV}},
  author = {Abelev, B. and others},
  collaboration = {ALICE},
  journal = {Phys. Rev. D},
  volume = {91},
  issue = {11},
  pages = {112012},
  numpages = {33},
  year = {2015},
  month = {Jun},
  publisher = {American Physical Society},
  doi = {10.1103/PhysRevD.91.112012},
  url = {https://link.aps.org/doi/10.1103/PhysRevD.91.112012}
}

@article{pythia6,
    author = "Sjostrand, Torbjorn and Mrenna, Stephen and Skands, Peter Z.",
    title = "{PYTHIA 6.4 Physics and Manual}",
    reportNumber = "FERMILAB-PUB-06-052-CD-T, LU-TP-06-13",
    doi = "10.1088/1126-6708/2006/05/026",
    journal = "JHEP",
    volume = "05",
    pages = "026",
    year = "2006"
}

@article{perugia,
  title = {Tuning Monte Carlo generators: The Perugia tunes},
  author = {Skands, Peter Z.},
  journal = {Phys. Rev. D},
  volume = {82},
  issue = {7},
  pages = {074018},
  numpages = {25},
  year = {2010},
  month = {Oct},
  publisher = {American Physical Society},
  doi = {10.1103/PhysRevD.82.074018},
  url = {https://link.aps.org/doi/10.1103/PhysRevD.82.074018}
}

@book{geant3,
      author        = "Brun, R and Bruyant, F and Maire, M and McPherson, A C and
                       Zanarini, P",
      title         = "{GEANT 3: user's guide Geant 3.10, Geant 3.11; rev.
                       version}",
      publisher     = "CERN",
      address       = "Geneva",
      year          = "1987",
      url           = "https://cds.cern.ch/record/1119728",
}

@article{DAlesio:2024ope,
    author = "D'Alesio, Umberto and Gamberg, Leonard and Murgia, Francesco and Zaccheddu, Marco",
    title = "{Transverse $\Lambda$ polarization in unpolarized $pp \to$ jet $\Lambda^{\uparrow}$ X}",
    reportNumber = "JLAB-THY-24-3993",
    doi = "10.1016/j.physletb.2024.138552",
    journal = "Phys. Lett. B",
    volume = "851",
    pages = "138552",
    year = "2024"
}

@article{Gao:2024bfp,
    author = "Gao, Ying and Chen, Kai-Bao and Song, Yu-Kun and Wei, Shu-Yi",
    title = "{Transverse polarization of Lambda hyperons in hadronic collisions}",
    doi = "10.1016/j.physletb.2024.139026",
    journal = "Phys. Lett. B",
    volume = "858",
    pages = "139026",
    year = "2024"
}

@article{Ji:2023cdh,
    author = "Ji, Zhaohuizi and Zhao, Xiao-Yan and Guo, Ai-Qiang and Xu, Qing-Hua and Zhang, Jin-Long",
    title = "{Lambda polarization at the Electron-ion collider in China}",
    primaryClass = "nucl-ex",
    doi = "10.1007/s41365-023-01317-w",
    journal = "Nucl. Sci. Tech.",
    volume = "34",
    number = "10",
    pages = "155",
    year = "2023"
}

@article{AbdulKhalek:2021gbh,
    author = "Abdul Khalek, R. and others",
    title = "{Science Requirements and Detector Concepts for the Electron-Ion Collider}: {EIC Yellow Report}",
    primaryClass = "physics.ins-det",
    reportNumber = "BNL-220990-2021-FORE, JLAB-PHY-21-3198, LA-UR-21-20953",
    doi = "10.1016/j.nuclphysa.2022.122447",
    journal = "Nucl. Phys. A",
    volume = "1026",
    pages = "122447",
    year = "2022"
}

@article{Boussarie:2023izj,
    author = "Boussarie, Renaud and others",
    title = "{TMD Handbook}",
    eprint = "2304.03302",
    archivePrefix = "arXiv",
    primaryClass = "hep-ph",
    reportNumber = "JLAB-THY-23-3780, LA-UR-21-20798, MIT-CTP/5386",
    month = "4",
    year = "2023"
}

\end{document}